\documentclass[runningheads]{llncs}
\usepackage[T1]{fontenc}
\usepackage{graphicx}
\usepackage{array}
\newcolumntype{P}[1]{>{\centering\arraybackslash}p{#1}}
\newcommand{\repeatthanks}{\textsuperscript{\thefootnote}}

\usepackage{soul,color}
\usepackage{float}
\usepackage{xurl}

\begin{document}
\title{Pixel-Level Explanation of Multiple Instance Learning Models in Biomedical Single Cell Images
}
\titlerunning{Pixel-Level Explanation of MIL}

\author{Ario Sadafi\inst{1,2}\thanks{equal contribution}
\and Oleksandra Adonkina\inst{1,3}\repeatthanks
\and Ashkan Khakzar \inst{2} 
\and \\ Peter Lienemann\inst{1}
\and Rudolf Matthias Hehr \inst{1}
\and Daniel Rueckert \inst{4}
\and \\ Nassir Navab \inst{2,5} 
\and Carsten Marr\inst{1}
\thanks{carsten.marr@helmholtz-munich.de}
}

\institute{Institute of AI for Health, Helmholtz Munich – German Research Center for Environmental Health, Neuherberg, Germany
\and
Computer Aided Medical Procedures (CAMP), Technical University of Munich, Germany
\and Faculty of Mathematics, Technical University Munich, Munich, Germany
\and
Artificial Intelligence in Healthcare and Medicine, Technical University of Munich, Munich,
Germany
\and 
Computer Aided Medical Procedures, Johns Hopkins University, USA
}

\authorrunning{A. Sadafi, O. Adonkina et al.}

\maketitle              % typeset the header of the contribution
\begin{abstract}
Explainability is a key requirement for computer-aided diagnosis systems in clinical decision-making. Multiple instance learning with attention pooling provides instance-level explainability, however for many clinical applications a deeper, pixel-level explanation is desirable, but missing so far. 
In this work, we investigate the use of four attribution methods to explain a multiple instance learning models: GradCAM, Layer-Wise Relevance Propagation (LRP), Information Bottleneck Attribution (IBA), and InputIBA. With this collection of methods, we can derive pixel-level explanations on for the task of diagnosing blood cancer from patients’ blood smears. We study two datasets of acute myeloid leukemia with over 100 000 single cell images and observe how each attribution method performs on the multiple instance learning architecture focusing on different properties of the white blood single cells. Additionally, we compare attribution maps with the annotations of a medical expert to see how the model's decision-making differs from the human standard.   
Our study addresses the challenge of implementing pixel-level explainability in multiple instance learning models and provides insights for clinicians to better understand and trust decisions from computer-aided diagnosis systems.
% positively impacting on the applicability of AI-based approaches.

\end{abstract}
\section{Introduction}
Healthcare systems are challenged by an increasing number of diagnostic requests and a shortage of medical experts. AI can alleviate this problem by providing powerful decision support systems that free medical experts from repetitive, tiring tasks \cite{rajpurkar2022ai}. However, explainability on all levels is required to ensure the proper working of deep learning 'black box' models, and to build trust for the widespread application of health AI \cite{khakzar2021towards,engstler2022interpretable}. 

 For decision making that relies on the analysis of hundreds of single instances (e.g. histological patches \cite{campanella2019clinical} or single cells \cite{sadafi2020attention}), attention-based multiple instance learning (MIL) provides explainability on the instance level \cite{aml}. This allows algorithms to highlight suspicious structures in cancer tissue
and retrieve prototypical, diagnostic cells in blood or bone marrow smears. In particular in cases where morphological features are unknown, it is of the highest importance to be able to inspect not only high attention instances, but at high attention pixels therein. 

A number of different approaches for pixel-level explainability have been proposed and evaluated in the past. 
Backpropagation based methods such as layer-wise relevance propagation (LRP) \cite{montavon2017explaining} and guided backpropagation \cite{springenberg2014striving} leverage the gradient as attribution. Other methods work with latent features, including GradCAM \cite{selvaraju2017grad}, which utilizes the activations on the final convolution layers, or IBA \cite{schulz2020restricting}, which measures the predictive information of latent features.
% - pixel level models: IBA, input IBA, LRP, .... explain what they do, how they differ. 
These methods are widely used in the medical field to provide some level of explainability: Böhle et at. \cite{bohle2019layer} use LRP to explain the decisions of the neural network on brain MRIs of Alzheimer disease patients; Arnaout et al. \cite{arnaout2021ensemble} propose an ensemble neural network to detect prenatal complex congenital heart disease and use GradCAM to explain the decisions of their expert-level model. IBA\cite{IBA} is explored in \cite{khakzar2021CheXIBA} for chest X-ray analysis. Another attribution method, InputIBA \cite{khakzar2021iba}, has proven to be useful for generating saliency maps for dermatology lesions \cite{krammer2022deep}. 

\begin{figure}[t]
\centering
\includegraphics[width=0.99\textwidth,page=1,trim=0cm 13cm 0.2cm 0cm,clip]{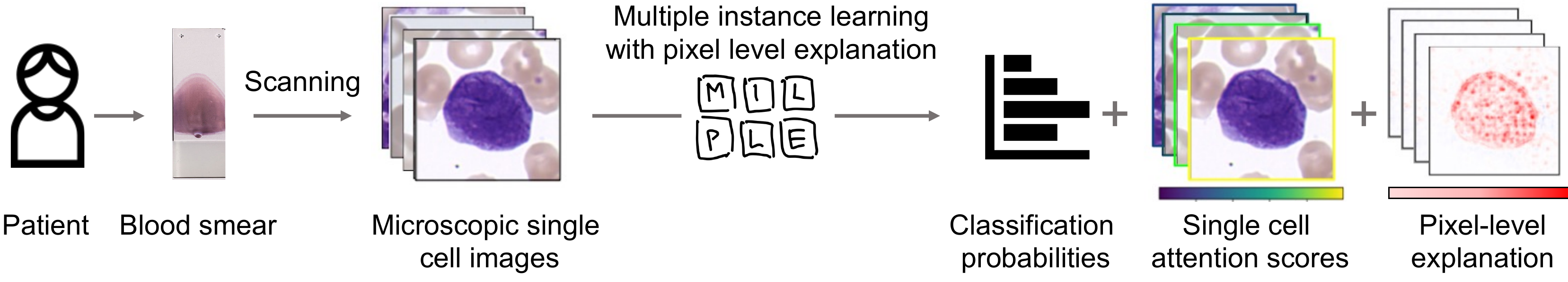}
\caption{MILPLE brings pixel-level explainability to multiple instance learning models. We apply MILPLE to two clinical single-cell datasets and showcase its explanatory power for revealing morpho-genetic correlations in blood cancer. In our example, blood smears from over 300 patients suffering from an aggressive leukemic subtype called acute myeloid leukemia (AML) have been digitized and microscopic images of white blood cells have been extracted. AML subtypes are predicted based on the pool of cells, and most important cells are identified based on the MIL attention mechanism, while the most important pixels in each of those are indicated with MILPLE.}
\label{figoverview}
\end{figure}

Unfortunately, most of these approaches cannot be applied to MIL out of the box. Complex gradient flows and the additional dimension introduced by the bag structure in the MIL model architecture requires adapting explainability algorithms accordingly. Here, we introduce MILPLE, the first multiple instance learning algorithm with pixel-level explainability. We showcase MILPLE (Fig. \ref{figoverview}) on two clinical single cell datasets with high relevance for the automatic classification of leukaemia subtypes from patient samples. 
We adapt GradCAM, LRP, IBA, and InputIBA to a MIL architecture and study the effectiveness of these methods in providing pixel-level explainability for instances. Although the quality of some of the methods seems visually plausible, quantitative analysis shows that there is no silver bullet addressing all challenges. With widespread applications of attention based MIL in different medical tasks, MILPLE helps provide pixel-level explanation using the mentioned algorithms. 
To foster reproducible research, our code is available on Github \url{https://github.com/marrlab/MILPLE}.
\section{Methodology}

\subsection{Multiple Instance Learning}

The objective of a multiple instance learning (MIL) model $f$ is analyzing a bag of input instances $B=\{I_1,...,I_N\}$ and classifying it into one of the classes $c_i \in C$ \cite{maron1997framework}. In attention-based MIL \cite{ilse2018attention}, an attention score $\alpha_k \in A$, $k \in \{ 1, ..., N\} $ for every instance quantifies the importance of that instance for bag classification:
\begin{equation}
    c_i, \alpha_k = f(B). 
\end{equation}
There are two approaches to implement MIL: Instance level and embedding level MIL \cite{vocaturo2019dangerousness}. We focus on the embedding level MIL, where every input instance is mapped into a low dimensional space via $h_k = f_\mathrm{emb}(I_k, \sigma)$ with $ \sigma$ being learned model parameter. By pooling information distributed between the instances, one bag is aggregated into a representative bag feature vector and used for the final classification. Attention pooling \cite{ilse2018attention} provides bag level of explainability and best accuracy in many problems.
MIL training can be formulated as
\begin{equation}
    \mathcal{L}_\mathrm{MIL}(\theta, \sigma) = \mathrm{CE}(c, \hat{c})
\end{equation}
with $\hat{c} = f_\mathrm{MIL}(H, A; \theta)$, where $c$ is the ground truth label for the whole bag, $H= \{h_1,...,h_N\}$ are the embedding feature vectors of all instances and CE is the cross entropy loss. $\theta$ and $\sigma$ represent learnable model parameters. Based on the attention scores $\alpha_k \in A$, the bag embedding $z$ is calculated as a weighted average over all of the embedding feature vectors:

\begin{equation}
    z = \sum_{k=1}^{N} \alpha_k h_k,
\quad\mathrm{where}\quad
    \alpha_k = \frac{\mathrm{exp}\{w^T \mathrm{tanh}(Vh_k^T)\}}{\sum^{N}_{j=1} \mathrm{exp}\{w^T \mathrm{tanh}(Vh_j^T)\}}.
\end{equation}
The parameters $V$ and $w$ are learned in a semi-supervised way during training. With only bag level annotation, instances with the most probable contribution to the classification are given a higher attention score. 
%The formulation of the attention ensures all values are within 0 and 1 and add up to 1. 
\subsection{GradCAM} 

Gradient-weighted Class Activation Mapping (GradCAM) is an explanation technique leveraging the gradient information to localize the most discriminative regions of an input image for a given model prediction. It computes the gradient of the predicted class score with respect to the feature maps of the last convolutional layer and weights each feature map by the corresponding gradient to obtain the class activation map. The class activation map highlights the regions of the input image that are most relevant for the prediction. Blue parts of the map indicate no contribution and red parts indicate high contribution.

\subsection{Layer-wise Relevance Propagation}
Layer-wise relevance propagation (LRP) is an explanation technique for deep neural networks which produces pixel-level decomposition of the input by redistributing relevance in the backward pass \cite{montavon2019layer}. Using local redistribution rules a relevance score $R_i$ is assigned to the input variable according to the classifier output $f(x)$:

\begin{equation}
    \sum_{i}R_i^0 = ... = \sum_{j}R_j^{L-2} = \sum_{k}R_i^{L-1} = ... = f(x)
\end{equation}

This backward distribution is lossless, meaning that no relevance is lost in the process while also no additional relevance is introduced at every layer $L$. A relevance score for every input variable $R_i$ shows the contribution of that variable to the final outcome, which is positive or negative, depending on whether that variable supported the outcome or went against the prediction.
The basic rule \cite{montavon2019layer} for LRP is defined as $R_j^{L-1}  = \sum_{k} \frac{a_j w_{jk}}{\sum_{j} a_j w_{jk}} R_k^{L}$, where $w_{jk}$ is the weight between the $j$ and $k$ layers and $a_j$ is the activation of neuron $j$. 
Eplison rule \cite{montavon2019layer} is an improvement to the basic rule by introducing a positive small $\epsilon$ value in the denominator.
%\begin{equation}
%        R_j^{L-1}  = \sum_{k} \frac{a_j w_{jk}}{\sum_{j} a_j w_{jk} + \epsilon} R_k^{L}
%\end{equation}
%where $w_{jk}$ is the weight between the $j$ and $k$ layers and $a_j$ is the activation of neuron $j$.
The $\epsilon$ will consume some of the relevance making sparser explanations with less noise.
Gamma rule \cite{montavon2019layer} tries to favor positive contributions more by introducing a $\gamma$ coefficient on positive weights 
%\begin{equation}
%    R_j^{L-1}  = \sum_{k} \frac{a_j (w_{jk} + \gamma w_{jk}^{+})}{\sum_{j} a_j (w_{jk} + \gamma w_{jk}^{+})} R_k^{L}
%\end{equation}
such that the impact on positive weights is controlled with it. As it increases, the effect of positive weights becomes more pronounced. 
ZBox rule \cite{montavon2017explaining} is designed for the input pixel space which is constraint to boxes. 
%\begin{equation}
%    R_j^{L-1} = \sum_{k}\frac{a_j w_{jk} - l_j w_{jk}^{-} - h_j w_{jk}^{+}}
%    {\sum_{j} a_j w_{jk} - l_j w_{jk}^{-} - h_j w_{jk}^{+}} R_k^{L}
%\end{equation}
%where $l_j$ and $h_j$ are the lowest and highest values possible for the pixels. 

\subsubsection{Application to MIL.}
MIL architectures are a complex combination of different layer types. Fully connected layers are more often used in earlier stages in comparison with normal convolutional neural networks. We tested different combinations of rules. Based on the results and suggestions introduced by Montavon et al. \cite{montavon2019layer}, we decided to apply ZBox rule on the first layer for every instance, gamma rule for the feature extractor $f_{\mathrm{emb}}$ and epsilon rule on the attention mechanism and final classifier.

\subsection{Information Bottleneck Attribution}
In contrast to LRP as a back-propagation method, Information bottleneck attribution (IBA) \cite{schulz2020restricting} is based on information theory. 
IBA works by placing a bottleneck on the network to restrict the flow of information by adding noise to the features. A bottleneck on the features $F$ at a given layer can be represented by $Z = \lambda F + (1-\lambda)\epsilon$ where $\epsilon$ is the noise controlled by $\lambda$, a mask with the same dimensions as F and elements with values between 0 and 1. 
The idea is to minimize the mutual information between the input $X$ and $Z$ while maximizing the information between $Z$ and target $Y$:
\begin{equation}\label{eq:information_bottleneck}
    \max_\lambda I(Y,Z) - \beta I(X,Z)
\end{equation}
Here, $\beta$ is the Lagrange multiplier controlling the amount of information that passes through the bottleneck.
$\mathcal{L}_I$ is an approximation of intractable term $I(X,Z)$:

\begin{equation}\label{eq:IBA_KL}
    I(X,Z) \approx \mathcal{L}_I = E_F[D_{KL}(P(Z|F)\parallel Q(Z)],
\end{equation}
where $Q(Z)$ is a normal distribution with estimated mean and variance of $F$ from a batch of samples. Intuitively, $I(Y,T)$ is equivalent to accurate predictions. Thus instead of maximizing it, we can minimize the loss function, cross entropy loss in our case, and therefore information bottleneck can be obtained by using $\mathcal{L} = \beta\mathcal{L}_{I} + CE$ as the objective.

\subsection{Input Information Bottleneck Attribution}
The motivation behind InputIBA \cite{khakzar2021iba} is to make the information bottleneck optimization in Eq. \ref{eq:information_bottleneck} possible on the input space. IBA as proposed in Eq. \ref{eq:information_bottleneck} and \ref{eq:IBA_KL} results in an overestimation of mutual information as the bottleneck is applied on earlier layers. The formulation is the most valid when the bottleneck is applied to a deep layer where the Gaussian distribution approximation of activation values is valid \cite{schulz2020restricting}.
Thus InputIBA proposes a trick where the optimal bottleneck is first computed using Eq. \ref{eq:information_bottleneck}. Let us refer to it as $Z^{*}$. Then we look for an input bottleneck $Z_{G}$ that induces the same optimal bottleneck on the deep layer. In order to make the input bottleneck $Z_{G}$ induce $Z^{*}$ in deep layers, the following distribution matching is done: 
\begin{equation}\label{eq:fit_dist}
    \min_{\lambda_{G}} D[P(f(Z_{G})) || P(Z^{*})]
\end{equation}
By optimizing Eq. \ref{eq:fit_dist} we find the optimal input bottleneck $Z_{G}^{*}$ that induces $Z^{*}$ in the selected deep layer. InputIBA proceeds to use $Z_{G}^{*}$ as a prior for solving the information bottleneck optimization (Eq. \ref{eq:information_bottleneck}). The input bottleneck $Z_{I}$ is conditioned on $Z_{G}$ as follows: $Z_{I} = \Lambda Z_{G} + (1-\Lambda)\epsilon$, where $\Lambda$ is the input mask. The final mask $Z_{I}^{*}$ is computed by optimizing Eq. \ref{eq:information_bottleneck} on $Z_{I}$, and it restricts the flow in the deep layers within limits defined by $Z_{G}^{*}$.

\subsubsection{Application to MIL.}
We had to overcome an obstacle of additional dimension introduced by the bag instances compared to conventional neural networks to apply InputIBA to the MIL structure. In comparison to standard neural networks working with single images, in MIL it is not straightforward to form a batch of bags as convolutions won't handle five dimensions. 
It is suggested to apply IBA on the deepest layer of the network, however in MIL architectures it seemed that applying IBA on earlier layers yields a better result. After conducting experiments and testing every convolution layers of the resnet backbone, we decided to place the bottleneck at the third convolutional layer where we obtained the best signal compared to other layers.
The distance in Eq. \ref{eq:fit_dist} is minimized based on an adversarial optimization scheme \cite{khakzar2021iba}. The generative adversarial network is trained for each instance in the bag individually. We used $\beta = 40$ to control the amount of information passing through the input bottleneck.

\subsection{Quantitative Evaluation of Pixel-wise Explainability Methods}
 There is extensive literature studying the quality of the explanations \cite{nie2018theoretical,adebayo2018sanity,sixt2020explanations,khakzar2022explanations,khakzar2021neural,khakzar2020rethinking,hooker2019benchmark}, but only few quantitative approaches exist. The intuition behind these methods is perturbation of features found to be important and measuring their impact on output to evaluate the quality of the feature attributions. \\
 \\
\textbf{Insertion/Deletion} \cite{samek2016evaluating}.
Insertion method gradually inserts pixels into the baseline input (zeros) while deletion method removes pixels from input data by replacing them with the baseline value (zero) according to their attribution scores from high to low.
While computing the output of the network over different percentage of insertion or deletion a curve is obtained. The area under the curve (AUC) is calculated for every input and averaged over the whole dataset.
A higher AUC in insertion means important pixels were inserted first while a lower AUC in deletion means important pixels were removed first.\\
\\
\textbf{Remove-and-Retrain} \cite{hooker2019benchmark}. (ROAR) is an empirical measure to approximate the quality of feature attributions by verifying the degradation of the accuracy of a retrained model when the features identified as important are removed from the dataset. The processes is repeated with various percentages of removal. A sharper degradation of the accuracy demonstrates a better identification of important features. Random assignment of importance is defined as a baseline.

\section{Experiments}

\subsection{Dataset}
We study the effectiveness of pixel attribution methods on acute meyleod leukimia (AML) subtype recognition tasks using two different datasets: DeepAPL and an in-house AML dataset. 

\noindent\textbf{DeepAPL} \cite{sidhom2021deep} is a single cell blood smear dataset consisting of 72 AML and 34 acute promyelocytic leukemia (APL) patients collected at the Johns Hopkins Hospital. 

\noindent \textbf{AML dataset} is a cohort of 242 patient blood smears from four different prevalent AML genetic subtypes \cite{khoury20225th}: i) APL with PML::RARA mutation, ii) AML with NPM1 mutation, iii) AML with RUNX1::RUNX1T1 mutation, and iv) AML with CBFB::MYH11 mutation. A fifth group of stem cell donors (SCD) comprises only healthy individuals and is thus used as the control group. 
Each blood smear contains at least 150 single white blood cell images resulting in a total of 81,214 cells. This dataset is available via TCIA\footnote{\url{https://doi.org/10.7937/6ppe-4020}}.

\begin{figure}[t]
\centering
\includegraphics[width=0.99\textwidth,page=2,trim=0cm 10.5cm 0cm 0cm,clip]{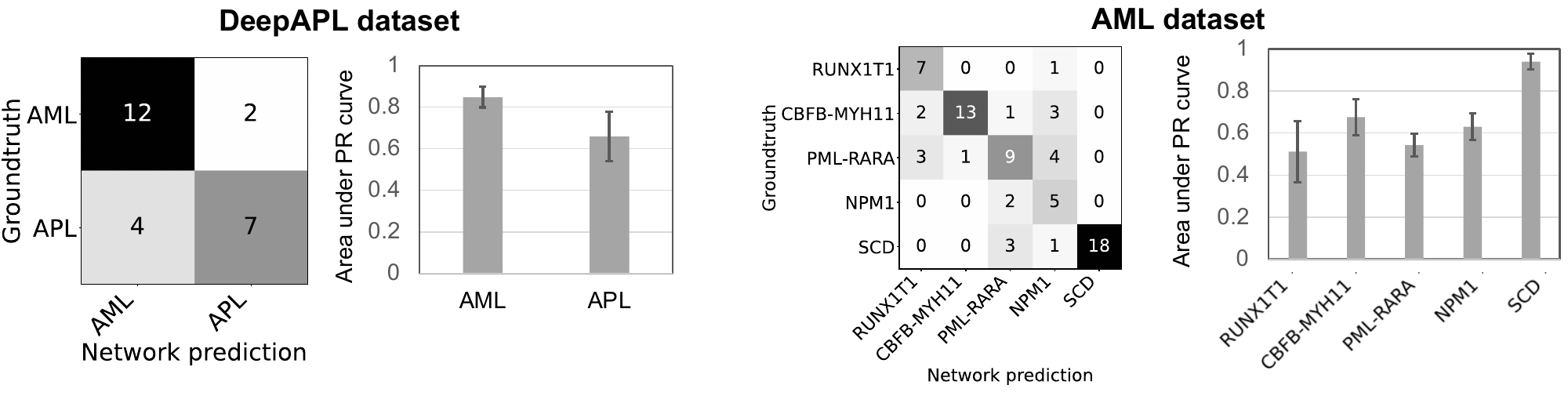}
\caption{The confusion matrix and area under the precision recall curve is reported for the two datasets MIL model was trained on. Mean and standard deviation are calculated over 5 independent runs.}
\label{figresults}
\end{figure}

\subsection{Implementation Details}
For the backbone of our approach and feature extraction from single cell images, we use the ResNeXt \cite{xie2017aggregated} architecture suggested by Matek et al. \cite{matek2019human}, which is pretrained on the relevant task of single white blood cell classification. Features are extracted from the last convolutional layer of the ResNeXt and passed into the MIL architecture with a second feature extraction step consisting of two convolutional layers with adaptive max-pooling and a fully connected layer. The attention mechanism consists of two fully connected layers and finally, the classifier consists of two fully connected layers. Adam Optimizer with a learning rate of $5\times10^4$ for DeepAPL and $5\times10^5$ for the AML dataset with a Nesterov momentum of $0.9$ was used. The datasets are split into stratified subsets for train, validation and test in a 60-20-20 percent regime.

\subsection{Model Training}
The training of the MIL model on the two datasets continues for 40 and 150 epoches,  respectively, while the validation loss is monitored. If the validation loss does not decrease for 5 consecutive epochs the training is stopped. We conducted 5 independent runs to train the model. Table \ref{tableresults} shows the mean and standard deviation of accuracy, macro F1 score and area under ROC curve.

\subsection{Evaluation of Explanations}
% We evaluate the quality of the explanations both qualitatively and quantitatively.

\subsubsection{Qualitative Evaluation} 
\begin{figure}[t]
\centering
\includegraphics[width=0.80\textwidth,page=5,trim=0cm 3cm 12cm 0cm,clip]{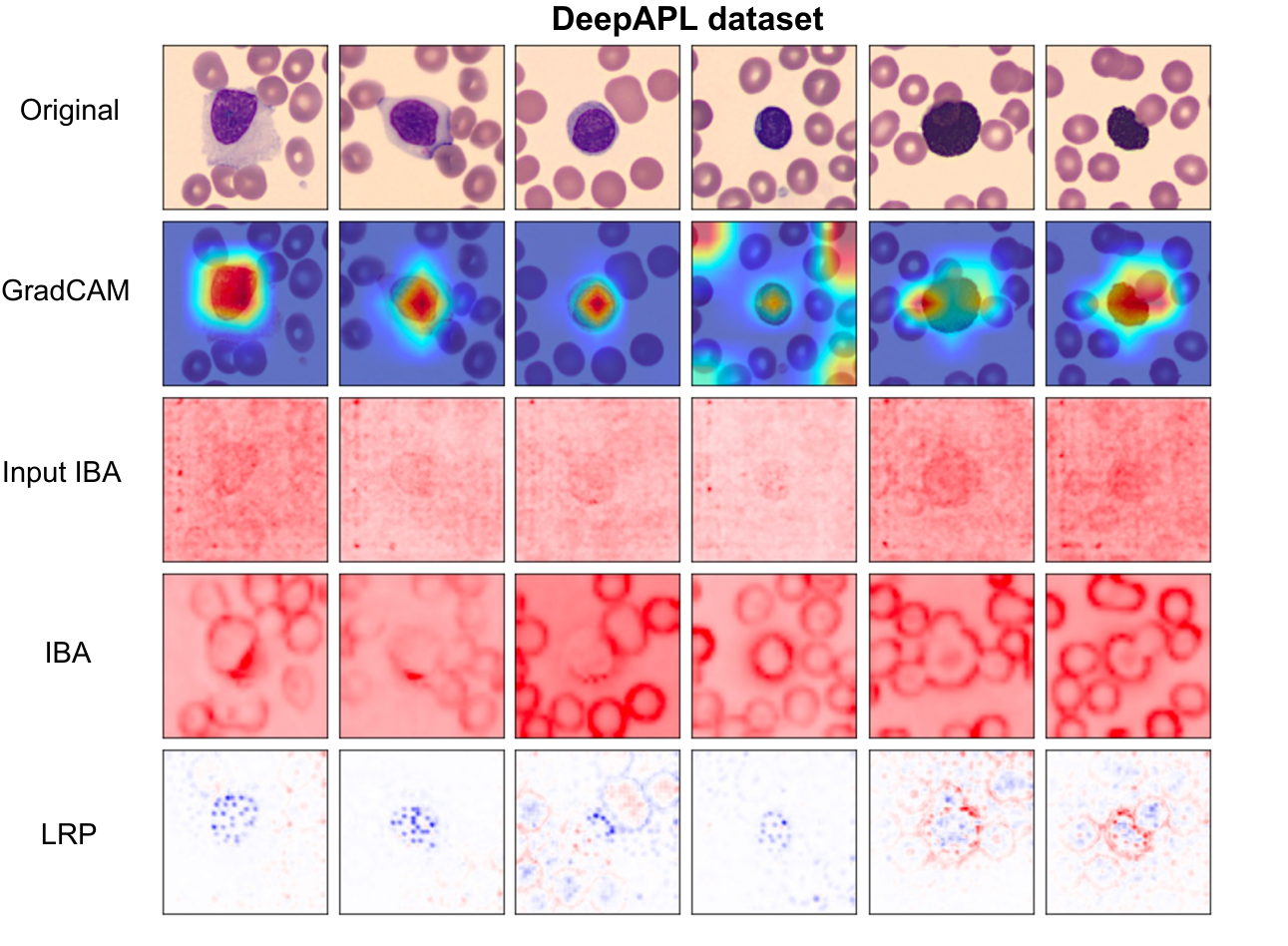}
\caption{Pixel-level explanation methods applied to exemplary images from the DeepAPL dataset. For GradCAM blue parts of the map indicate no contribution and red parts indicate high contribution, and similarly, for LRP  blue parts indicate negative contribution and red parts indicate positive contribution. In many cases GradCAM and LRP focus on the white blood cells in the center of the image, while IBA focuses also on the red blood cell surrounding it. InputIBA shows a relatively scattered focus.} 
\label{figvisual1}
\end{figure}

\begin{figure}[ht]
\centering
\includegraphics[width=0.85\textwidth,page=3,trim=0cm 2cm 11.7cm 0cm,clip]{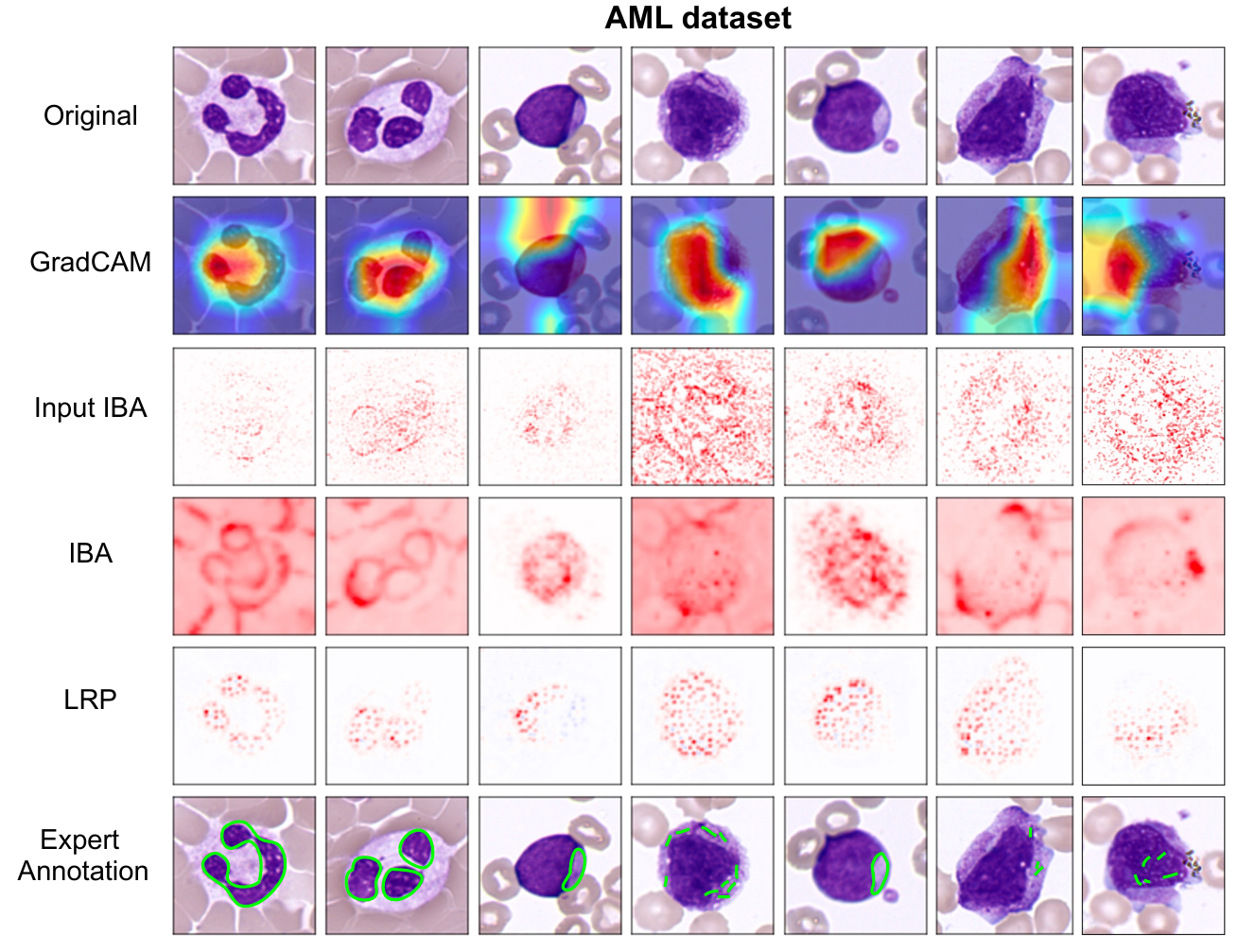}
\caption{Pixel-level explanation methods applied on exemplary images from AML dataset. In the first two images, all methods agree on the morphology found relevant by the expert (last row). In the following images, the methods highlight different regions and are only sometimes in concordance with the expert.} 
% LRP provides highly accurate, finely detailed pixel wise explanation for cells in a multiple instance learning architecture. Comparing different pixel attribution methods applied on the example cells from the bag, we compare GradCAM, Input IBA, IBA and LRP. Complex and weak gradient flow leads to inaccurate pixel attributions for most of the methods.}
\label{figvisual2}
\end{figure}

includes inspection of single cell images and comparison with medical expert annotation. 
Fig. \ref{figvisual1} and \ref{figvisual2} show selected cells from both datasets and pixel-level explanations provided by the four different methods. In Fig. \ref{figvisual2}, we compare pixel attributions with expert annotations as a medical expert has annotated a small subset of single cells in the AML dataset. Most of the methods detect morphological features defined by the expert as important. 

% Qualitatively, LRP seems the most probable approach for pixel-wise explanation for multiple instance learning architectures. 


\subsubsection{Quantitative Evaluation} 
\begin{figure}[t]
\centering
\includegraphics[width=0.95\textwidth,page=4,trim=0cm 8.7cm 4cm 0cm,clip]{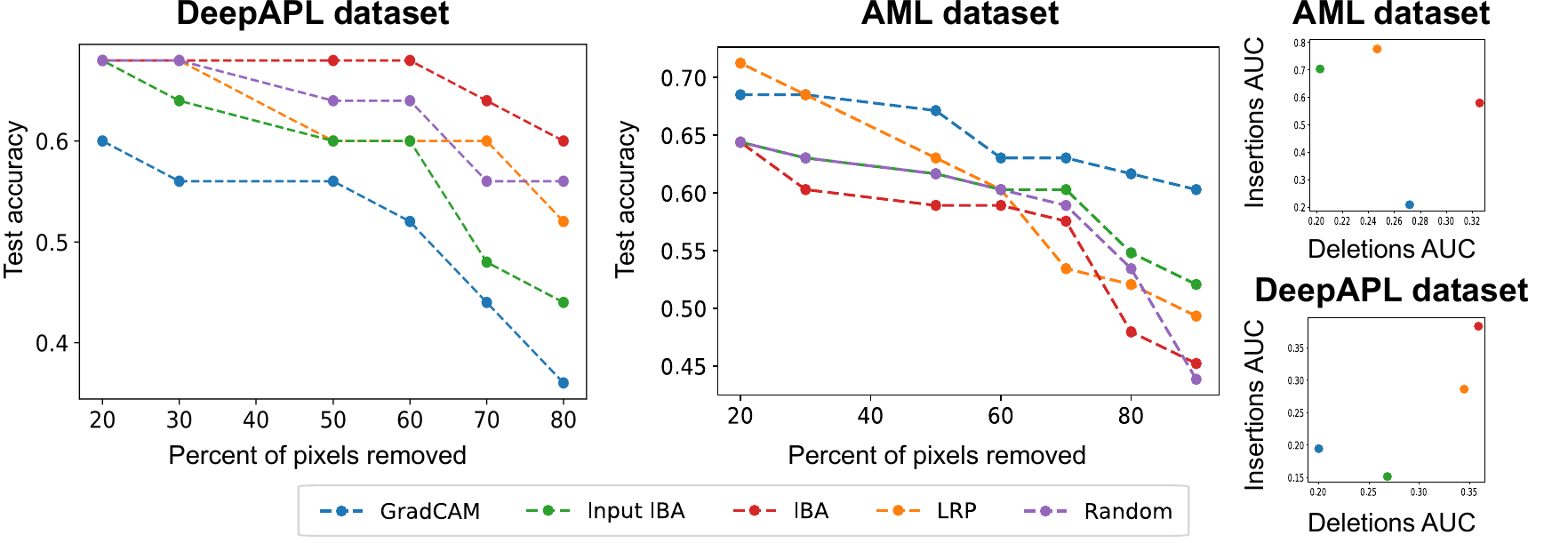}
\caption{Remove-and-Retrain (ROAR) experiment (left) and insertion/deletion experiment (right) for both datasets. GradCAM has the best pixel attribution in DeepAPL ROAR experiments, while on the AML dataset, LRP and IBA perform best. Insertion/deletion experiments for DeepAPL support GradCAM. For the AML dataset InputIBA and LRP have the best performance in insertion/deletion experiment. }
\label{figroar}
\end{figure}

of the explanations is an essential step for correct understanding of what model focuses on. In order to evaluate the quality of different methods, we performed Insertion/Deletion and ROAR experiments on each of the GradCAM, LRP, IBA, and InputIBA methods as shown in Fig. \ref{figroar}. The performance of the method is highly dependent on the dataset and each time different methods end up to be the most suitable

\subsection{Discussion and Results}
\textbf{Model Performance:} We compare our training on DeepAPL with the state-of-the-art method proposed by Sidhom et al. \cite{sidhom2021deep} on the dataset. Since the datasets are imbalanced, we are reporting the area under precision recall curve for each class as well as the confusion matrix for both datasets to get a better view over the class-wise performance. Fig. \ref{figresults} shows that classification results are robust across the two datasets. On DeepAPL, with no special tailoring of the method to the dataset, we could outperform the state-of-the-art method based on sample analysis cell by cell. MIL takes all cells into consideration and can thus achieve a higher accuracy in the task. On the AML dataset some ambiguity exists between different malignant classes, which is to be expected since AML subtype classification based on cell morphology only is a challenging task even for the medical experts. Model identifies the majority of benign stem cell donors correctly.\\
\\
\textbf{Explanations:} A close inspection of the pixel explanations from the four different methods reveals fundamental differences (see Fig. \ref{figvisual1},\ref{figvisual2}):
For the DeepAPL dataset (Fig. \ref{figvisual1}) we observe that GradCAM focuses on the white blood cell nucleus in most cases. In some cells however it fails to recognise the cell and instead puts high relevance on background pixels at the image border. Though according to our ROAR results (Fig. \ref{figroar}), removing the white blood cell affects the accuracy significantly, pointing to the fact that the network is using features relevant to them. InputIBA puts most focus on the centre of the image, and thus correctly on the white blood cell. However, pixel attention is spread out over the whole image at times (Fig. \ref{figvisual1}). The ROAR results for InputIBA (Fig. \ref{figroar}) also show that the accuracy drops if corresponding image regions are removed.  

\begin{table}[t]
\caption{Mean and standard deviation of accuracy, macro F1 score and area under ROC curve is reported for the two blood cancer datasets. Our attention based MIL method outperforms the original DeepAPL method \cite{sidhom2021deep}.}
\label{tableresults}
\begin{center}
\begin{tabular}{p{1.5cm}|P{2.5cm}|P{2.5cm}|P{2.5cm}|P{2.5cm}}
\textbf{Data} &\textbf{Method} & \textbf{Accuracy} & \textbf{F1 score} & \textbf{AU ROC} \\\hline
DeepAPL &  ours  & $0.65 \pm 0.07$ & $0.63 \pm 0.08$ & $0.750 \pm 0.078$    \\\hline
DeepAPL  & Sidhom et al. \cite{sidhom2021deep} & - & - & $0.739$  \\\hline\hline
AML & ours & $0.68 \pm 0.03$ & $0.65 \pm 0.04$ & $0.855 \pm 0.037$ \\

\end{tabular}
\end{center}
\end{table}

On the AML dataset we observe that IBA highlights image regions that correspond to either abnormal cytoplasm (4th, 6th and 7th cell from left, Fig. \ref{figvisual2}) or to structures in the nucleus (first two cells in Fig. \ref{figvisual2}). These are particularly interesting since they show that the method is able to retrieve morphological details that escape the human eye (3rd cell: the cell appears to be dark violet in the original images, but IBA is able to focus on morphology therein) and to segment granules, whose structure is relevant for cell type classification (4th cell). The ROAR results from the AML dataset (Fig. \ref{figroar}) show that removing morphological features identified by IBA significantly disrupts accuracy. This signifies that the model relies on these pixel during training.
LRP focuses on the white blood cell in the image center and the nucleus therein. We observe that the ROAR results for LRP are not very informative (Fig. \ref{figroar}), and the method performs similarly to random. This might be due to the LRP structure and a problem with the ROAR metric. However, LRP achieves a good score on the Deletion/Insertion metric (Fig. \ref{figroar}). This means that LRP features have an immediate effect on the output of the network.

\section{Conclusion}
Incorporating pixel-level explainability in multiple instance learning allows us to inspect instances, evaluate the focus of our model, and find morphological details that might be missed by the human eye. All four pixel-level explainability methods we used revealed interesting insights and highlighting morphological details that fit prior expert knowledge. 
% For the single cell data we analysed here, IBA seems to be most relevant for highlighting morphological details that fit to prior knowledge on morpho-genetic correlations. 
%
However, more work has to be done on systematically comparing and quantifying clinical expert annotations with explainability predictions, to eventually select appropriate methods for the application at hand, and potentially reveal novel morpho-genetic correlations. 

We believe that our study will be instrumental for multiple instance learning applications in health AI. Single-cell data is ideal for method development, since it allows a direct comparison of model prediction and human intuition. However, applied to computational histopathology, where a large amount of digitized data exists, the pixel-level insight into tissue structure at multiple scales might reveal morphological properties previously unrecognized. With novel spatial single-cell RNA sequencing technologies being on the brink of becoming available widely, we expect a high demand for methods like MILPLE.

\bibliographystyle{splncs04}
\bibliography{article}
\end{document}